\documentclass[a4paper,11pt]{article}
\pdfoutput=1

\usepackage{jheppub}

\usepackage{multirow}

\usepackage{subfig}
\usepackage[countmax]{subfloat}

\usepackage{comment}

\usepackage{color}
\usepackage{graphicx}                   
\usepackage{bbm}
\usepackage{amsmath}
\usepackage{amssymb}
\usepackage{bbold}
\usepackage{xspace}
\usepackage{verbatim}
\usepackage{bm}

\newcommand{\mathsym}[1]{{}}
\newcommand{\unicode}[1]{{}}

\newcommand{\nbar}{{\bar n}}

\DeclareRobustCommand{\App}[1]{App.~\ref{#1}}

\DeclareRobustCommand{\Fig}[1]{Fig.~\ref{#1}}

\DeclareRobustCommand{\Eq}[1]{Eq.~(\ref{#1})}

\DeclareRobustCommand{\Ref}[1]{Ref.~\cite{#1}}
\DeclareRobustCommand{\Refs}[1]{Refs.~\cite{#1}}

\preprint{ \begin{flushright}
LA-UR-16-27653
\end{flushright}
}

\title{The Asymptotic Form of Non-Global Logarithms, Black Disc Saturation, and Gluonic Deserts}
\author[*]{Duff Neill}
\affiliation[*]{Theoretical Division, MS B283, Los Alamos National Laboratory, Los Alamos, NM 87545, USA}

\emailAdd{duff.neill@gmail.com}

\abstract{We develop an asymptotic perturbation theory for the large logarithmic behavior of the non-linear integro-differential equation describing the soft correlations of QCD jet measurements, the Banfi-Marchesini-Smye (BMS) equation. This equation captures the late-time evolution of radiating color dipoles after a hard collision. This allows us to prove that at large values of the control variable (the non-global logarithm, a function of the infra-red energy scales associated with distinct hard jets in an event), the distribution has a gaussian tail. We compute the decay width analytically, giving a closed form expression, and find it to be jet geometry independent, up to the number of legs of the dipole in the active jet. Enabling the asymptotic expansion is the correct perturbative seed, where we perturb around an anzats encoding formally no real emissions, an intuition motivated by the buffer region found in jet dynamics. This must be supplemented with the correct application of the BFKL approximation to the BMS equation in collinear limits. Comparing to the asymptotics of the conformally related evolution equation encountered in small-x physics, the Balitisky-Kovchegov (BK) equation, we find that the asymptotic form of the non-global logarithms directly maps to the black-disc unitarity limit of the BK equation, despite the contrasting physical pictures. Indeed, we recover the equations of saturation physics in the \emph{final} state dynamics of QCD. 
}

\begin{document} 
\maketitle
\section{Introduction}
QCD jet cross-sections with distinct measurements in different phase space regions present a fundamental difficulty in their perturbative description. Soft QCD emissions can populate all angular regions, and cause distinct jet measurements to become correlated. Fundamentally, different jet regions are not independent factorized objects for naive jet measurements. These soft correlations are embodied in the jet cross-section in the form of so-called non-global logarithms (NGLs), first described in the seminal work of \Ref{Dasgupta:2001sh}, which are logarithms of the ratio of infra-red energy scales associated with distinct jets. Much effort has been expended on analyzing NGLs, from their expansion at fixed order \cite{Kelley:2011tj,Hornig:2011iu,Hornig:2011tg,Kelley:2011aa,Kelley:2012kj,Schwartz:2014wha,Khelifa-Kerfa:2015mma,Delenda:2015tbo}, to numerical resummations and phenomenology\cite{Dasgupta:2002dc,Dasgupta:2002bw,Appleby:2002ke,Rubin:2010fc,Banfi:2010pa,Hatta:2013iba,Hagiwara:2015bia,Larkoski:2015npa}, to developing the theory of their evolution equations \cite{Banfi:2002hw,Weigert:2003mm,Marchesini:2003nh,Hatta:2008st,Avsar:2009yb,Caron-Huot:2015bja,Becher:2015hka,Becher:2016mmh,Caron-Huot:2016tzz}, as well as their relationship to soft jet factorization theorems \cite{Larkoski:2015zka,Larkoski:2015kga,Neill:2015nya,Larkoski:2016zzc}. The classical example is back-to-back dijets resulting from an $e^+e^-$ collision, where each jet has a hemispherical region associated with it. One can impose distinct mass cuts in each hemisphere, and if $m_H$ is the mass scale associated with the heavier hemisphere, and $m_L$ is the mass scale of the lighter hemisphere, the cross-section will be a function of the logarithm:\footnote{In general, we will let $m_H$ and $m_L$ refer to the mass scales of the active-jet region and the softer jet region respectively.}
\begin{align}\label{eq:NGL_def}
L&=\frac{C_A}{\pi}\int_{m_L}^{m_H}\frac{d\mu}{\mu}\alpha_s(\mu)\approx\frac{\alpha_s C_A}{\pi}\text{ln}\frac{m_H}{m_L}\,.
\end{align}
If we demand that $m_H,m_L\ll Q$, where $Q$ is the center of mass energy of the collision, than we have a genuine dijet event, and logarithms of $Q$ and the jet masses (the \emph{global} logarithms of the cross-section) that are associated with the hard production of the initial jets can be resummed via the renormalization group evolution of a factorization theorem, as formulated in soft-collinear effective field theory (SCET) \cite{Bauer:2000ew, Bauer:2000yr, Bauer:2001ct, Bauer:2001yt,Bauer:2002nz,Stewart:2009yx,Hornig:2009vb,Ellis:2010rwa,Stewart:2010tn}, or through the coherent branching formalism of \Refs{Catani:1990rr,Catani:1992ua,Dokshitzer:1998kz,Banfi:2004yd,Banfi:2014sua}. Since there are no real emissions at large angles between the hard scale $Q$ and the heavy jet scale $m_H$, the resummation of these global logarithms has a simple exponential form, at least in some conjugate space, and for a wide variety of jet shape observables. 

Non-global logarithms in contrast depend sensitively on the secondary branching history in the more energetic jet regions. Formally, all emissions in the energetic region must be tracked before a single soft emission sets the scale of the softest regions. In the multi-color limit of QCD, a master equation that describes the branching history has been developed, called the Banfi-Marchesini-Smye (BMS) equation \cite{Banfi:2002hw}, whose leading logarithmic form is:
\begin{align}\label{eq:BMS_LL}
\partial_L g_{ab}&=\int_J\frac{d\Omega_j}{4\pi}W_{ab}(j)\Big(U_{abj}g_{aj}g_{jb}-g_{ab}\Big)\,.
\end{align}
$g_{ab}$ is the NGL distribution associated with an initial color dipole with null directions $a=(1,\hat{a})$ and $b=(1,\hat{b})$.\footnote{At leading logarithmic level for the hemisphere observable, $g_{ab}$ can be taken to be the cumulative cross-section for the hemisphere masses, divided by the square root of the product the cumulative cross-sections for the total hemisphere jet mass at $m_L$ and at $m_H$. In the laplace conjugate space, this definition works to all orders as the definition of a purely non-global observable.} We will interchangeably refer to points on the celestial sphere surrounding the hard scattering by both angular coordinates, their unit vector, or the associated null direction.  The active jet region is donated by $J$, and this is where real emissions can populate between the scales $m_H$ and $m_L$, and thus is the integration region for the emission $j$. The nonlinear term describes the production of real gluons in a strongly ordered (in energies) limit, splitting the initial dipole into two new ones, and the linear term is the virtual correction that renders the equation infra-red and collinear (IRC) safe. The production of these real emissions is given by $W_{ab}(j)$, the leading-order eikonal factor for soft emissions:
\begin{align}\label{eq:LO_eikonal_factor}
W_{ab}(j)&=\frac{a\cdot b}{a\cdot j\,j\cdot b}\,,\\
a\cdot b&= 1-\hat{a}\cdot\hat{b}\,.
\end{align}
The $U_{abj}$ factor is a resummation factor that results from factoring out all global logarithms from the soft distribution, so that the equation describes purely non-global correlations. It has the form:
\begin{align}\label{eq:U_factor}
U_{abj}&=\text{Exp}\Big[L\int_{\bar{J}}\frac{d\Omega_q}{4\pi}\Big(W_{aj}(q)+W_{jb}(q)-W_{ab}(q)\Big)\Big]\,.
\end{align}
The integration is over the complement to the active jet region $J$, denoted $\bar{J}=S^2-J$. The color dipole function $g_{ab}$ must satisfy the following initial and boundary conditions:
\begin{align}
g_{ab}(0)&=1\,,\\
\lim_{a\parallel b}g_{ab}&=1,\qquad\forall L\,.
\end{align}
The former is simply the statement that we start with two emitters after the hard collision, and the second is necessary to guarantee the collinear safety of \Eq{eq:BMS_LL}. The hard evolution kernel of the active jet region for the BMS equation is known with full color dependence to two loops in QCD, and in the large-$N_c$ limit to three loops in $N=4$ super-yang-mills (SYM) theory, see \Refs{Caron-Huot:2015bja,Caron-Huot:2016tzz}. The BMS equation can also be derived from factorization theorems describing soft jet production, where the resummation factor $U$ arises naturally from the renormalization group evolution of these factorization theorems, as described in \Refs{Larkoski:2015zka,Neill:2015nya}. This interpretation leads naturally to a systematic approximation to the full BMS solution in terms of a sum over a finite number of resummed soft jets, and in \Ref{Larkoski:2016zzc} it was proved that such a series has an infinite radius of convergence.\footnote{For a related expansion used in the context of rapidity gaps in hadron collisions, see \Refs{Forshaw:2006fk,Forshaw:2008cq,Forshaw:2009fz,DuranDelgado:2011tp}.} Moreover, examining the analytic structure of these dressed gluons revealed an essential limitation to the fixed order description of the NGLs, where it was shown that the series had a finite radius of convergence. This is due to the inability of fixed-order perturbation theory to describe the emergent buffer region of \Ref{Dasgupta:2002bw}. At large values of the NGLs, real emissions are dynamically suppressed near the edge of the active jet region, leading to an asymptotic desert of real emissions except for collinear regions about the initial hard legs.

Due to the fact that the BMS equation is a non-linear integro-differential equation, analytic results about its solutions have been limited, and mainly confined to small $L$ regions. Though the BMS equation is collinear safe, this is due to a cancellation between the real and virtual corrections. The numerical methods of \Refs{Dasgupta:2001sh,Hatta:2013iba,Schwartz:2014wha,Hagiwara:2015bia} therefore implement an explicit regularization of the collinear divergences, either through a lattice regularization or an explicit cutoff in the smallest angular size of a dipole. In the far-tail region, one becomes highly sensitive to this collinear cutoff. In this paper we give a recipe for an asymptotic perturbation theory for the full BMS equation at large $L$, where we exploit properties of the collinear regulated BMS equation introduced in \Ref{Larkoski:2016zzc}. This regulated equation admits an analytically tractable linearization, and we show how to construct a collinear regulator that asymptotes to the true BMS distribution. Thus we derive the leading asymptotic form of the NGL distribution at large-$N_c$, showing it to be Gaussian and giving an closed form expression for its decay width. An essential role is played by the buffer region in this asymptotic analysis. The perturbative seed for the large $L$ behavior supposes that the NGL distribution is dominated by the virtual correction in the almost all the active jet region except for a small collinear region about the initial hard jets. That is, there are no real emissions are large angles.

The outline of the paper is as follows. First we introduce the collinearly regulated BMS equation, and its linearization. We then derive the asymptotic condition the collinear cutoff must obey in order for the linearization to asymptote to the true distribution. We then focus on the back-to-back initial dipole case, with a conical active jet region around one of the legs. Here we can work completely analytically, and derive the Gaussian form and the decay width, with the Balitsky-Fadin-Kuraev-Lipatov (BFKL) equation \cite{Kuraev:1977fs,Balitsky:1978ic} playing a key role at small angles.  We then comment on the relationship of these results to the related Balitsky-Kovchegov (BK) equation \cite{Balitsky:1995ub, Kovchegov:1999yj} describing unitarization and saturation effects in small-x resummations of the parton distribution function, finding a direct mapping to the solution governing black disc unitarity. We then conclude.

\section{Bounding the Tail}
We investigate how to use the collinearly regulated BMS equation to derive the true asymptotic form of the NGL distribution. Our plan of attack is as follows:
\begin{itemize}
\item Deform the BMS equation into an explicitly collinearly regulated version.
\item Form a linear differential inequality for the collinearly regulated BMS equation by dropping the non-linear terms.
\item The differential inequality is easily solved for an arbitrary $L$-dependent cutoff.
\item Use this solution to the differential inequality for the collinearly regulated BMS equation as a seed for an asymptotic expansion for the full BMS equation.
\item At small angles inside the buffer region, make use of the BFKL equation to approximate the splittings of the full BMS equation.
\item Derive the necessary functional form of the cutoff such that the solution to the differential inequality is genuinely asymptotic to the full solution of the BMS equation.
\end{itemize}
That such a lower bound could describe the genuine NGL tail relies on the buffer region dynamics. As the NGL becomes larger, emissions must cluster about the initial hard legs. Physically, one is vetoing collinear splittings at the edge of the jet. The regions away from these legs, but still in the active jet region, become an asymptotic desert. That is, there is no important soft jet production in those regions, and only the virtual correction matters. 

To regulate the BMS equation, we make use of the so-called $\delta$-regulator of the eikonal lines, see \Refs{Chiu:2009yx,GarciaEchevarria:2011rb,Echevarria:2015usa,Echevarria:2015byo}, where we add a small constant to the eikonal propagators. In \App{sec:eikonal_integrals}, we describe the behavior of the regulated eikonal factor using the $\delta$-regulator. Alternatively, one could use an explicit hard cutoff on the angular integrations, as was done in \Ref{Larkoski:2016zzc} in proving the existence of solutions to the BMS equation via the dressed gluon expansion. Importantly, the small angle behavior and large angle behavior, where the angle is the angle between the legs of the eikonal factor, is reproduced with both regularizations at small cutoffs, however, the $\delta$-regulatization procedure is more analytically tractable. We write the regulated BMS equation:
\begin{align}
\partial_Lg_{ab}^{\delta}&=\int_{J}\frac{d\Omega_j}{4\pi}W_{ab}^{\delta}(j)\Big(U_{abj}(L)g_{aj}^{\delta}g_{jb}^{\delta}-g_{ab}^{\delta}\Big)\,,\\
W^{\delta}_{ab}(j)&=\frac{a\cdot b}{(a\cdot j+\delta^2)(j\cdot b+\delta^2)}\label{eq:cut_off_eikonal}\,.
\end{align}
$g^{\delta}$ is the solution of the collinearly regulated BMS equation. We can now define the regulated BMS kernel acting in a region $J$ on a function $g$ as:
\begin{align}
G_{ab}^{\delta}[g;J]&=\int_{J}\frac{d\Omega_j}{4\pi}W_{ab}^{\delta}(j)\Big(U_{abj}(L)g_{aj}g_{jb}-g_{ab}\Big)\,.\label{eq:BMS_kernel_in_region}
\end{align}
We adopt the convention that $G$ and $W$ with no superscripts are simply the zero regulator limits that enter into the full BMS equation:
\begin{align}
W_{ab}(j)=W_{ab}^{\delta=0}(j)& &G_{ab}[g;J]=G_{ab}^{\delta=0}[g;J]
\end{align}
It is a simple matter to note that solutions to the collinearly regulated BMS equation are bounded from below by integrating over the virtual correction alone. That is, we can drop the nonlinear terms in the regulated integral to form a differential inequality:
\begin{align}\label{eq:diff_inequality}
\partial_Lg_{ab}^{\delta}&\geq-\Bigg(\int_{J}\frac{d\Omega_j}{4\pi}W_{ab}^{\delta}(j)\Bigg)g_{ab}^{\delta}=-\gamma_{ab}(\delta)g_{ab}^{\delta}
\end{align}
This follow from the fact that the real emission terms are everywhere positive in the active jet region. If we convert the differential inequality into an equality, then solutions to this new differential equation will bound from below solutions to the regulated BMS equation when they share identical initial conditions. This is called the comparison theorem for differential inequalities, see \Ref{CorduneanuDiff}. Thus we introduce $g_{ab}^{\gamma(\delta)}$, the solution to the inequality for the initial condition $g^{\delta}_{ab}(0)=1$:
\begin{align}\label{eq:Col_Reg_Solutions_Bounded}
g_{ab}^{\delta}&\geq g_{ab}^{\gamma(\delta)}\\\label{eq:Col_Reg_Solutions_Bounded_Form}
g_{ab}^{\gamma(\delta)}&=\text{Exp}\Big[-\int_0^{L}d\ell\,\gamma_{ab}\Big(\delta(\ell)\Big)\Big]
\end{align}
Note that we have allowed the regulator to depend on the $L$. We can feed the bounding solution in \Eq{eq:Col_Reg_Solutions_Bounded} into the \emph{unregulated} BMS equation. Then we want to find the $\delta$ such that the bounding solution asymptotically solves the BMS equation:
\begin{align}\label{eq:asymptotic_condition}
\lim_{L\rightarrow \infty}\Big(g^{\gamma(\delta)}_{ab}\Big)^{-1}G_{ab}\Big[g^{\gamma(\delta)}; J\Big]+\gamma_{ab}\Big(\delta(L)\Big)=0
\end{align}
If this goes to zero as $L\rightarrow\infty$, we conclude that the lower bound on the collinearly regulated BMS equation is approximating the full solution, that is:
\begin{align}
\lim_{L\rightarrow\infty}\frac{g_{ab}}{g^{\gamma(\delta)}_{ab}}\rightarrow 1
\end{align}
Of course, this is simply the statement that the collinearly regulated lower bound is asymptotic to the full solution. The essential simplification in \Eq{eq:asymptotic_condition} is that for a given collinear cutoff, we have a explicit analytic form to perturb around. Thus we can derive an expression for the full BMS kernel acting upon the collinearly regulated lower bound that is asymptotically accurate. For back-to-back initial jets, we will find that the kernel can be asymptotically expanded and analytically evaluated as $L\rightarrow\infty$. For general dipoles, we will find that \Eq{eq:asymptotic_condition} can only be satisfied when the leading asymptotic form of the NGL distribution is Gaussian, and derive an explicit expression for the decay width. However, we must bear in mind the buffer region is only active at large angles, and at the smallest dipole opening angles, we will find that the collinearly regulated BMS equation does not approximate the full solution, even asymptotically. This will be checked by expanding and solving the BMS equation in collinear limits, where it will reduce to the BFKL equation. Thus at small angles, we will use the solution to the BFKL equation to integrate the BMS kernel in the asymptotic condition \eqref{eq:asymptotic_condition}. Demanding the large angle integration using the collinear cutoff ansatz is consistent with the BFKL solution then fixes the asymptotic form.

\section{Back-to-Back Jets}\label{sec:back_to_back}
To simplify our lives, we focus on the back-to-back case with conical jets, so that $a=n=(1,0,0,1)$ and $b=\nbar=(1,0,0,-1)$, but with a cone around the active jet region of radius $R$. We then have the BMS kernel acting on the collinearly regulated lower bound:
{\small\begin{align}\label{eq:BMS_lower_bound_btob_error}
\Big(g^{\gamma(\delta)}_{n\nbar}\Big)^{-1}G_{n\nbar}\Big[g^{\gamma(\delta)}; J\Big]=\int_{0}^{R}\frac{d\theta}{\text{sin}\,\theta}\Bigg(\text{Exp}\Big[L\,\text{ln}\Big(1-\frac{\text{tan}^2\frac{\theta}{2}}{\text{tan}^2\frac{R}{2}}\Big)-\int_{0}^{L}d\ell\Big(\gamma_{nj}(\delta(\ell))+\gamma_{j\nbar}(\delta(\ell))-\gamma_{n\nbar}(\delta(\ell))\Big)\Big]-1\Bigg)\,.
\end{align}}
Now $\theta$ is the angle of the emission $j$ to the jet axis $\hat{n}$, and we have used the azimuthal symmetry of the back-to-back case to do the azimuthal integral. Each regulated eikonal factor in the numerator, $\gamma_{jn}$ or $\gamma_{j\nbar}$, is simply a function of $\theta$. 

\section{Calculation at Large Angles}\label{sec:asymptotics}
In this section we derive the asymptotic form of the BMS kernel acting on the ansatz \eqref{eq:Col_Reg_Solutions_Bounded_Form} for a generic cutoff that is exponentially decreasing as $L\rightarrow\infty$, as in \Eq{eq:general_cutoff}. By asymptotic, we mean strictly that the ratio of either side of an equation is 1 as $L\rightarrow\infty$.

The action of the BMS kernel on the collinearly regulated bound in \eqref{eq:BMS_lower_bound_btob_error} is still too complicated to integrate analytically. However, the terms in the exponent of the kernel can be simplified further by expanding in the collinear limit $\theta\rightarrow 0$.  The important observation is (see \App{sec:eikonal_integrals}) that when $\delta\ll 1$:
\begin{align}\label{eq:angular_behavior_of_soft_eikonal}
0\gg-\text{ln}\Big(1+\frac{\text{sin}^2\frac{\theta}{2}}{\delta^2(\ell)}\Big)\geq-\gamma_{nj}(\delta(\ell))-\gamma_{j\nbar}(\delta(\ell))+\gamma_{n\nbar}(\delta(\ell))\geq -\frac{\theta^2}{4\delta^2(\ell)}\,,\forall \theta
\end{align}
In the strict collinear limit this is becomes a $\approx$ for both the upper and lower bounds:
\begin{align}\label{eq:angular_behavior_of_soft_eikonal_small_angles}
-\gamma_{nj}(\delta(\ell))-\gamma_{j\nbar}(\delta(\ell))+\gamma_{n\nbar}(\delta(\ell))= -\frac{\theta^2}{4\delta^2(\ell)}+O\Big(\frac{\theta^4}{\delta^4},\theta^2,\delta^2\Big)
\end{align}
Thus we introduce the following integrals:
\begin{align}\label{eq:BMS_lower_bound_btob_error}
\mathcal{G}^<_{n\nbar}(\delta;R)&=\int_{0}^{R}\frac{d\theta}{\text{sin}\theta}\Bigg(\text{Exp}\Big[L\,\text{ln}\Big(1-\frac{\text{tan}^2\frac{\theta}{2}}{\text{tan}^2\frac{R}{2}}\Big)-\frac{\theta^2}{4}\int_{0}^{L}\frac{d\ell}{\delta^2(\ell)}\Big]-1\Bigg)\nonumber\\
\mathcal{G}^>_{n\nbar}(\delta;R)&=\int_{0}^{R}\frac{d\theta}{\text{sin}\theta}\Bigg(\text{Exp}\Big[L\,\text{ln}\Big(1-\frac{\text{tan}^2\frac{\theta}{2}}{\text{tan}^2\frac{R}{2}}\Big)-\int_{0}^{L}d\ell\text{ln}\Big(1+\frac{\text{sin}^2\frac{\theta}{2}}{\delta^2(\ell)}\Big)\Big]-1\Bigg)
\end{align}
We then have:
\begin{align}
\mathcal{G}^>_{n\nbar}(\delta;R)\geq\Big(g^{\gamma(\delta)}_{n\nbar}\Big)^{-1}G_{n\nbar}\Big[g^{\gamma(\delta)}; J\Big]\geq\mathcal{G}^<_{n\nbar}(\delta;R)
\end{align}
Taking an exponential form of the cutoff:
\begin{align}\label{eq:general_cutoff}
\delta(L)&=\delta_i  e^{-\frac{L}{\kappa}}\,,
\end{align}
we can do the $\ell$ integrations in the exponent analytically, and then it is a simple matter to verify numerically that:
\begin{align}
\lim_{L\rightarrow\infty}\frac{\mathcal{G}^<_{n\nbar}(\delta;R)}{\mathcal{G}^>_{n\nbar}(\delta;R)}=1
\end{align}
This is true for any given fixed $R,\kappa$. This is not surprising, since both integrals are dominated by the behavior at $\theta=0$, and all the integrals have the same limits in that region. For the lower bound, we can derive an analytic form that is valid up to exponentially small corrections as $L\rightarrow\infty$. The genuine kernel is then also asymptotic to the lower bound, being sqeezed between the two limits. For the lower bound, terms at large $\theta$ (away from the collinear limit) in the exponential terms of \eqref{eq:BMS_lower_bound_btob_error} are exponentially small, and thus give exponentially small contributions to the integral. Replacing the terms in exponent with their collinear expansion in \Eq{eq:angular_behavior_of_soft_eikonal} throughout the $\theta$ integration range will then result in an exponentially small error, vanishing as $L\rightarrow\infty$.  We write:
\begin{align}
\mathcal{G}^>_{n\nbar}(\delta;R)&=\int_{0}^{R}\frac{d\theta}{\theta}\Bigg(\text{Exp}\Big[-k(L)\theta^2\Big]-1\Bigg)-\int_{0}^{R}d\theta\Big(\frac{1}{\text{sin}\theta}-\frac{1}{\theta}\Big)+...\\
k(L)&=\frac{1}{4}\int_0^{L}\frac{d\ell}{\delta^2(\ell)}=\frac{\kappa}{2}\delta^{-2}(L)+...
\end{align}
We can evaluate the first integral using the definitions of incomplete gamma functions, $\Gamma[\nu,x]$, and the second by expanding the inverse sine, integrating term by term, and summing the series:
\begin{align}
\mathcal{G}^>_{n\nbar}(\delta;R)&=-\frac{1}{2}\Bigg(\gamma_E+\Gamma\big[0,k(L)R^2\big]+\text{ln}\Big(k(L)R^2\Big)\Bigg)-\sum_{n=1}^{\infty}\frac{2(-1)^{n+1}(2^{2n-1}-1)B_{2n}}{2n(2n)!}R^{2n}+...\\
&=-\frac{1}{2}\Bigg(\gamma_E+\Gamma\big[0,k(L)R^2\big]+\text{ln}\Big(2k(L)\text{tan}^2\frac{R}{2}\Big)\Bigg)+...
\end{align}
We can even let $R$ be a function of $L$, and as long as:
\begin{align}
\lim_{L\rightarrow\infty}k(L)R^2(L)\rightarrow\infty\,,
\end{align}
then we have the result:
\begin{align}\label{eq:leading_order_asymptotics_I}
\Big(g^{\gamma(\delta)}_{n\nbar}\Big)^{-1}G_{n\nbar}\Big[g^{\gamma(\delta)}; J\Big]=-\frac{1}{2}\Bigg(\gamma_E+\text{ln}\Big(2k(L)\text{tan}^2\frac{R}{2}\Big)\Bigg)+...\,,
\end{align}
where we have dropped all terms subleading in the asymptotic limit $L\rightarrow\infty$. This holds true even for logarithmic growth to infinity of $kR^2$. In particular, if we take the difference between the kernel acting in conical jet regions about the direction $n$ of two different radii $r<R$, we have the asymptotic result:\footnote{$D_a^r$ is defined in \App{sec:eikonal_integrals}}
\begin{align}\label{eq:leading_order_asymptotics_II}
\Big(g^{\gamma(\delta)}_{n\nbar}\Big)^{-1}G_{n\nbar}\Big[g^{\gamma(\delta)}; D_n^R\Big]-\Big(g^{\gamma(\delta)}_{n\nbar}\Big)^{-1}G_{n\nbar}\Big[g^{\gamma(\delta)}; D_n^r\Big]&=-\int_r^{R}\frac{d\theta}{\sin\theta}+...&=-\text{ln}\frac{\text{tan}\frac{R}{2}}{\text{tan}\frac{r}{2}}+...
\end{align}
That is, we would have gotten the same result if we simply chopped out the real emission terms, and integrated the virtual contribution between the two radii.

\section{The BFKL Equation and its Solution}
Before turning to the NGL distribution for dipoles at large NGL values but small opening angles, we first introduce some notation and results about the position space BFKL equation. The BFKL equation in position space for color dipoles can be written as (see \Ref{Mueller:1994gb,Mueller:1994jq}):
\begin{align}
\label{eq:BFKL_transverse_space}\partial_t \Phi_{ab}(t)&=\int_{\mathbb{R}^2}\frac{d^2\vec{x}_j}{2\pi} \frac{\vec{x}_{ab}^2}{\vec{x}_{aj}^2\vec{x}_{jb}^2}\Big(\Phi_{aj}(t)+\Phi_{jb}(t)-\Phi_{ab}(t)\Big)\\
\vec{x}^2_{ab}&=(\vec{x}_a-\vec{x}_b)^2
\end{align}
The solution to the BFKL equation can be written in terms the eigenfunctions of the kernel, which have the form:
\begin{align}\label{eq:BFKL_eigen}
\chi(\nu)\Phi_{ab}^{\nu}&=\int_{\mathbb{R}^2}\frac{d^2\vec{x}_j}{2\pi} \frac{\vec{x}_{ab}^2}{\vec{x}_{aj}^2\vec{x}_{jb}^2}\Big(\Phi_{aj}^{\nu}+\Phi_{jb}^{\nu}-\Phi_{ab}^{\nu}\Big)+...\\
\chi(\nu)&=-2\gamma_E-\psi(0,1-\nu)-\psi(0,\nu)\label{eq:BFKL_characteristic}\\
\Phi_{ab}^{\nu}&\propto(\vec{x}_{ab}^{2})^{\nu}
\end{align}
$\psi(0,\nu)$ is the 0-th poly-gamma function, and from the characteristic function $\chi(\nu)$, we see that we have poles at integer $\nu$. In general we restrict:
\begin{align}
0<\nu<1
\end{align} 
The BFKL equation is used to resum the logarithms of center-of-mass energy $s$ over the momentum transfer $t$ in the partonic description of the cross-section for forward scattering of two highly boosted QCD composite states fired at each other. To use the position space BFKL equation, we take the composite states to be a collection of color dipoles, where each leg of the dipole is an eikonalized gluon located at a specific point in the transverse plane to the boost direction. The size of the dipole is the euclidean distance in the transverse plane between the two eikonal legs in the dipole. The BFKL equation describes the production of secondary soft radiation of the color dipoles composing the QCD state as a function of boost of that state. That is, if a dipole is given an extra boost, the BFKL equation describes how the dipole can radiate a soft gluon, under the assumption that the initial conditions of the composite state are ``dilute,'' that is, only a limited number of small color dipoles carry the total momentum of the composite state. As we will see in the next section, by taking the appropriate collinear limit of the BMS equation, we will recover the BFKL equation. An arbitrary solution to the BFKL equation can be represented as a superposition of the eigenfunctions: 
\begin{align}\label{eq:BFKL_solutions}
\Phi_{ab}(t)&=\int_{\nu_0-i\infty}^{\nu_0+i\infty}\frac{d\nu}{2\pi i}A({\nu})e^{t\chi(\nu)}(\vec{x}_{ab}^{2})^{\nu}
\end{align}
As we will show in the next section, the BFKL equation will also describe the production of secondary soft radiation within small color dipoles, where now we consider the dipole legs to be distributed in the celestial sphere around a hard scattering event, as opposed to the transverse plane. By small, we mean that the opening angle of the two eikonal lines is taken to be small. The BFKL equation will describe how soft emissions occur in such small dipoles at large NGL values.

\section{BFKL at Small Angles}
One could be tempted to take the action of the BMS kernel on the ansatz found in \Eq{eq:leading_order_asymptotics_I} as the final answer, and attempt to solve \Eq{eq:asymptotic_condition} directly using the anzats in all angular integration regions. The issue with the result in \Eq{eq:leading_order_asymptotics_I} is the treatment of the small angle region when using the ansatz in \Eq{eq:Col_Reg_Solutions_Bounded_Form} everywhere. In the back-to-back case of \Eq{eq:BMS_lower_bound_btob_error}, this corresponds to when $j\parallel n$. To see that the ansatz fails for small dipoles, we go back to the BMS equation at large $L$, but at small angles. Now we can expand the equation in the collinear limit, where $a\parallel b$. In the collinear limit, the solutions obey $g_{ab}\sim 1$, so we write \Eq{eq:BMS_LL} as:
\begin{align}
g_{ab}&=1+\Phi_{ab}+...\\
\label{eq:BFKL}\partial_L \Phi_{ab}&=\int_{0}^{\infty}\theta_j d\theta_j\int_{0}^{2\pi}\frac{ d\phi_j}{2\pi} \frac{\theta_{ab}^2}{\theta_{aj}^2\theta_{jb}^2}\Big(\Phi_{aj}+\Phi_{jb}-\Phi_{ab}\Big)+...
\end{align}
We have adopted a method-of-regions point of view \cite{Beneke:1997zp}, that to get the asymptotic behavior at small angles, we should expand everything in the collinear limit, including the phase space limits of integration. Indeed, we have also expanded the equation as if $j\parallel a,b$.\footnote{We have dropped terms originating from $U_{abj}$ : as we will see, this is self-consistent, once we examine solutions to \Eq{eq:BFKL_solution}. We have $U_{ abj} = 1 + O(\theta_{ab}^2) + ..$ which is power suppressed relative to the BFKL solution.} As discussed in \Ref{Marchesini:2003nh}, \Eq{eq:BFKL} is nothing but the BFKL equation in position space for color dipoles as given above in \Eq{eq:BFKL_transverse_space}. Indeed, after expanding the solution $g_{ab}$ and the $U$ factor, in the collinear limit, and taken the phase space to the whole celestial sphere, we could have simply conformally mapped the equation to the plane, leaving the eikonal factor alone. Now from Eqs. \eqref{eq:BFKL_eigen} and \eqref{eq:BFKL_solutions}, we take the NGL distribution at small dipoles to be given by:
\begin{align}\label{eq:BFKL_solution}
g_{ab}^{\gamma(\delta)}&\rightarrow 1+A\Big(\frac{\theta_{ab}}{\theta_c(L)}\Big)^{2\nu}+...\\
\label{eq:cutoff_from_BFKL}\theta_c(L)&=\text{Exp}\Big(-\frac{\chi(\nu)}{2\nu}L\Big)
\end{align} 
We could attempt a more sophisticated solution of the BFKL equation, however, such a simple form is sufficient to derive the slowest possible asymptotic decay of the NGL distribution. When we examine the ansatz \eqref{eq:Col_Reg_Solutions_Bounded_Form} at small angles using \Eq{eq:angular_behavior_of_soft_eikonal_small_angles}, we find we \emph{do} have the form of a BFKL eigenfunction, but with an infinite eigenvalue! Thus the ansatz is not correct at small angles, and we should not be using \Eq{eq:angular_behavior_of_soft_eikonal} at the smallest angles of integration of the kernel in the condition \eqref{eq:asymptotic_condition}. Physically, at the smallest angles but large values of the logarithm, we are \emph{inside} the buffer region, where $\theta_c(L)$ gives the angle between the initial hard legs and the boundary of the buffer region as a function of the NGL. Here soft emissions are unrestricted. The anzats from the collinearly regulated BMS equation formally encodes no emissions, and so it is not surprising the solution must be modified in the small angle limit. 

\section{BMS Asymptotic Condition Accounting for BFKL}
\begin{figure}\centering
\includegraphics[scale=.3]{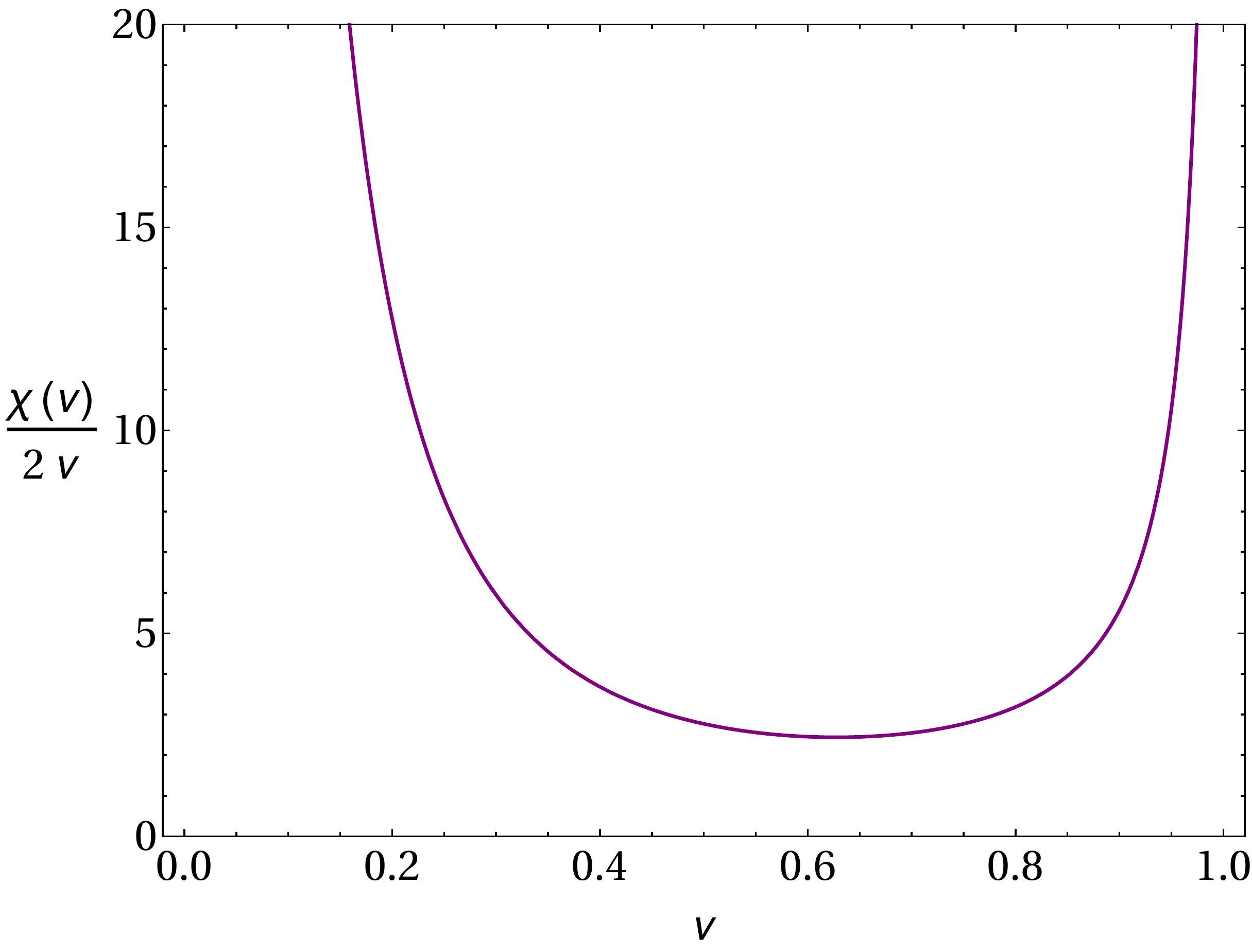}
\caption{\label{fig:chi} The coefficient of the NGL in the exponent of the critical angle. We plot as a function of the parameter $\nu$, angular scaling exponent for the solution to the BFKL equation found in the collinear limit. The true asymptotics is taken to be the minimum.}
\end{figure}

So we now return to \Eq{eq:asymptotic_condition}. In the action of the BMS kernel, we split the angular integration into two regions, one at small angles where we use the BFKL solution of \Eq{eq:BFKL_solution}, and the asymptotic result in \Eq{eq:leading_order_asymptotics_II} at large angles. That is, we write:
{\small\begin{align}\label{eq:BMS_kernel_on_ansatz_BFKL_improved}
&\Big(g^{\gamma(\delta)}_{n\nbar}\Big)^{-1}G_{n\nbar}\Big[g^{\gamma(\delta)}; J\Big]+\gamma_{n\nbar}(\delta;J)\nonumber\\
&\qquad=\Big(g^{\gamma(\delta)}_{n\nbar}\Big)^{-1}G_{n\nbar}\Big[g^{\gamma(\delta)}; D_n^{\theta_c}\Big]+\Bigg(\Big(g^{\gamma(\delta)}_{n\nbar}\Big)^{-1}G_{n\nbar}\Big[g^{\gamma(\delta)}; D_n^{R}\Big]-\Big(g^{\gamma(\delta)}_{n\nbar}\Big)^{-1}G_{n\nbar}\Big[g^{\gamma(\delta)}; D_n^{\theta_c}\Big]\Bigg)+\gamma_{n\nbar}(\delta;J)
\end{align}}
The first term will be evaluated using the BFKL solution, the second term will be evaluated using the large angle anzats. Large and small angles are defined relative to when we leave the region of validity for the BFKL solution, $\theta_c$ in \Eq{eq:cutoff_from_BFKL}. That is, when the opening angle of the color dipole is of order the critical angle, we no longer have the condition $g_{ab}\sim 1$. At small angles we have:
{\begin{align}
\Big(g^{\gamma(\delta)}_{n\nbar}\Big)^{-1}G_{n\nbar}\Big[g^{\gamma(\delta)}; D_n^{\theta_c}\Big]&=\int_0^{\theta_c}\frac{d\theta}{\text{sin}\theta}\Big(U_{n\nbar j}\frac{g_{jn}^{BFKL}g_{j\nbar}^{\gamma(\delta)}}{g_{n\nbar}^{\gamma(\delta)}}-1\Big)\approx A\int_0^{\theta_c}\frac{d\theta}{\theta}\Big(\frac{\theta}{\theta_c(L)}\Big)^{2\nu}+...
\end{align}}
Where we made use of \Eq{eq:BFKL_solution}. Then to find the full asymptotic condition, we add in the large angle contribution, and using \Eq{eq:nnbar_gamma} for the $\gamma_{n\nbar}$, giving the result:
{\small\begin{align}\label{eq:BMS_kernel_on_ansatz_BFKL_improved}
&\lim_{L\rightarrow\infty}\gamma_{n\nbar}(\delta;J)+\Big(g^{\gamma(\delta)}_{n\nbar}\Big)^{-1}G_{n\nbar}\Big[g^{\gamma(\delta)}; J\Big]\nonumber\\
&\qquad\qquad=\gamma_{n\nbar}(\delta;J)-\int_{\theta_c}^{R}\frac{d\theta}{\text{sin }\theta}+ \frac{A}{\theta_c^{2\nu}}\int_{0}^{\theta_c}\frac{d\theta}{\theta}\theta^{2\nu}+...\\
&\qquad\qquad=-\text{ln}\frac{\delta(L)}{\sqrt{2}\text{tan}\frac{R}{2}}+\text{ln}\frac{\text{tan}\frac{\theta_c}{2}}{\text{tan}\frac{R}{2}}+\frac{A}{2\nu}+...\\
&\qquad\qquad=-\text{ln}\delta(L)-\frac{L\chi(\nu)}{2\nu}+...
\end{align}}
Importantly, all geomentry dependence $R$ has dropped, and the constants can always be tuned with $\delta_i$ or equivalently $A$ to give zero, and contribute subleading terms to the leading behavior. Therefore:
\begin{align}\label{eq:asymptotic_cutoff}
\delta(L)&=\delta_i\text{exp}\Big(-\frac{\chi(\nu)}{2\nu}L\Big)\\
\label{eq:asymptotic_BMS}\text{ln} g_{n\nbar}&=-\frac{\chi(\nu)}{4\nu}L^2+...
\end{align}
The smallest value that $\frac{\chi(\nu)}{4\nu}$ can attain, and hence the slowest asymptotics the NGL distribution could possibly have is:
\begin{align}
\nu_{min}&\approx 0.62755\label{eq:crit_nu_NGL}\\
\frac{\chi(\nu_{min})}{4\nu_{min}}&\approx 1.2208\label{eq:gaussian_coef_NGL}
\end{align}
This result we take to be true for arbitrary dipoles in arbitrary jet regions. The collinear limit of the BMS equation loses all dependence on jet geometry. When adding in the large angle contribution given in \Eq{eq:leading_order_asymptotics_II}, it is also dominated by the collinear poles. Since it is given by integrating over the virtual correction over the same jet regions, it will always exactly match the geometry dependence of the ansatz in \Eq{eq:Col_Reg_Solutions_Bounded_Form}, up to subleading details of the differing regularization. It is a simple matter to show that when both legs of the dipole are in the active jet region, the decay is twice as strong. So we have the general results:
\begin{align}
\label{eq:asymptotic_BMS_gen_in-in}\text{ln} g_{ab}\Big|_{\text{in-in}}&=-\frac{\chi(\nu_{min})}{2\nu_{min}}L^2+...\\
\label{eq:asymptotic_BMS_gen_in-ou}\text{ln} g_{ab}\Big|_{\text{in-out}}&=-\frac{\chi(\nu_{min})}{4\nu_{min}}L^2+...
\end{align}
These results represent the slowest possible asymptotics, and there is no reason at this level of analysis to suppose that the NGL distribution will decay faster, though technically one could envision that we must have a very specific solution in the collinear region to the BFKL equation, demanding faster asymptotics. More generally, we could see the parameter $\nu$ as indexing the possible family of asymptotic solutions, and we have picked the stationary (minimum) point of this family as representing the true asymptotics. In \Fig{fig:chi}, we plot the dependence on the decay width on the parameter $\nu$.

\section{Comparison to the Asymptotics of the BK equation}
The BMS equation is conformally equivalent to the BK equation when $J=S^2$ and $U=1$. We then have the two equations side by side:
\begin{align}\label{eq:BMS_BK}
\partial_Lg_{ab}&=\int_{S^2}\frac{d\Omega_j}{4\pi}W_{ab}(j)\Big(g_{aj}g_{jb}-g_{ab}\Big)\nonumber\\
\partial_{Y}S(\vec{x}_a,\vec{x}_b)&=\frac{\alpha_sC_A}{2\pi^2}\int_{\mathbb{R}^2}d^2\vec{x}_j\frac{\vec{x}^2_{ab}}{(\vec{x}_{aj})^2(\vec{x}_{jb})^2}\Big(S(\vec{x}_a,\vec{x}_j)S(\vec{x}_j,\vec{x}_a)-S(\vec{x}_a,\vec{x}_b)\Big)
\end{align}
where we define:
\begin{align}
\vec{x}^2_{ab}&=(\vec{x}_a-\vec{x}_b)^2\,.
\end{align}
$S$ is interpreted as the forward scattering $S$-matrix. There are several different regimes of approximate solutions to the BK equation, however, the Gaussian decay of the NGL distribution suggests we focus upon the black-disc limit. When examining the Levin-Tuchin asymptotic solution \cite{Levin:1999mw}, valid in the unitary limit where the QCD target nucleus is a black disk, one first imposes a cutoff on the integral over the transverse plane, set by the saturation scale $Q_s(Y)$:
\begin{align}
\int_{\mathbb{R}^2}d^2\vec{x}_j\rightarrow \int_{\mathbb{R}^2}d^2\vec{x}_j\theta\Big((\vec{x}_a-\vec{x}_j)^2-\frac{1}{Q_s^2(Y)}\Big)\theta\Big((\vec{x}_b-\vec{x}_j)^2-\frac{1}{Q_s^2(Y)}\Big)
\end{align}
$Y$ is the rapidity of the boosted dipole with respect to the beam axis, and the rapidity dependent saturation scale is given as \cite{Triantafyllopoulos:2002nz,Mueller:2002zm,Mueller:2004sea,Iancu:2004es}:
\begin{align}\label{eq:saturation_scale}
Q_s(Y)&=Q_{si}\text{Exp}\Big(\frac{\alpha_{s}C_A}{\pi}\frac{\chi(\nu_{crit})}{\nu_{crit}}Y-\frac{3}{2\nu_{crit}}\text{ln}\Big(\frac{\alpha_{s}C_A}{\pi}Y\Big)+...\Big)\\
\chi(\nu)&=-2\gamma_E-\psi(\nu)-\psi(1-\nu)
\end{align}
$\nu_{crit}$ is given in \Eq{eq:crit_nu_NGL}. With the cutoff imposed, then the real emission terms can be dropped, in a manner very similar to the formation of the differential inequality for the collinearly regulated BMS equation,\footnote{However, the use of the comparison theorem for differential equations requires one pays careful attention to the initial conditions. The initial conditions of BMS evolution do not obviously map to that of a QCD bound state.} so that:
\begin{align}
\partial_{Y}S(\vec{x}_a,\vec{x}_b)&\sim-\frac{\alpha_sC_A}{\pi}\gamma_{\vec{x}_a\vec{x}_b}(Q_s^{-1}(Y))S(\vec{x}_a,\vec{x}_b)\\
S(\vec{x}_a,\vec{x}_b)&=S_0\text{Exp}\Big(-\frac{\alpha_sC_A}{\pi}\int_{Y_i}^{Y_f}\gamma_{\vec{x}_a\vec{x}_b}(Q_s^{-1}(Y))\Big)\label{eq:levin_tuchin_start}
\end{align}
In a slight abuse of notation, we denote the integration over the transverse propagators by the same $\gamma$ as used for BMS case, bearing in mind the cutoffs break the conformal relations between the two objects. Nevertheless, the collinear singularities are \emph{universal}:
\begin{align}\label{eq:BK_levin_tuchin_solution}
\frac{\alpha_sC_A}{\pi}\gamma_{\vec{x}_a\vec{x}_b}(Q_s^{-1}(Y))&\propto\frac{\alpha_sC_A}{\pi}\text{ln}\Big(\vec{x}_{ab}^2Q_s^2(Y)\Big)+...
\end{align}
This almost conformally related to \Eq{eq:regulated_eikonal_limits}, when $\theta_{ab}<r=\pi$, once we make the identification of the collinear regulator with the saturation scale:
\begin{align}\label{eq:collinear_regulator_to_saturation}
\delta^2(L)\leftrightarrow Q_s^{-1}(Y)
\end{align}
The angular function does not quite map directly to the transverse distance, but rather we have the correspondence under the conformal mapping:
\begin{align}
\frac{a\cdot b}{2-a\cdot b}&=\text{tan}^2\frac{\theta_{ab}}{2}\leftrightarrow x_{ab}^2
\end{align}
We have shown the relation in \Eq{eq:collinear_regulator_to_saturation} to be valid to leading order in the asymptotic expansion. 

What is particularly fascinating is that the form of the saturation scale given in \Eq{eq:saturation_scale} has been argued to be a consequence of universal properties of branching random walks \cite{Munier:2003vc, Munier:2003sj, Iancu:2004es}. Indeed, under this interpretation, one is tempted to posit that the collinear cutoff marks the boundary of the effective size of the buffer region and the region where active emissions can still occur, much like a boundary between the saturated region of the random walk and the leading edge. Paradoxically, NGL evolution is in some sense this process in reverse: the buffer region is where no emissions occur, and the front of this region, measured relative to the hard initial legs of the dipole, is \emph{decreasing} with the increasing NGL. Moreover, if this correspondence can be made precise, then the subleading asymptotic behavior of the collinear cutoff will take the same form as the saturation scale of \Eq{eq:saturation_scale}, since these subleading orders are controlled by the average position of such fronts within branching random walks \cite{Brunet:1997zz,1999CoPhC.121..376B,2000PhyD..146....1E,Brunet:2005bz}.

\section{Numerical Comparison To Monte Carlo}
\begin{figure}\centering
\includegraphics[scale=.3]{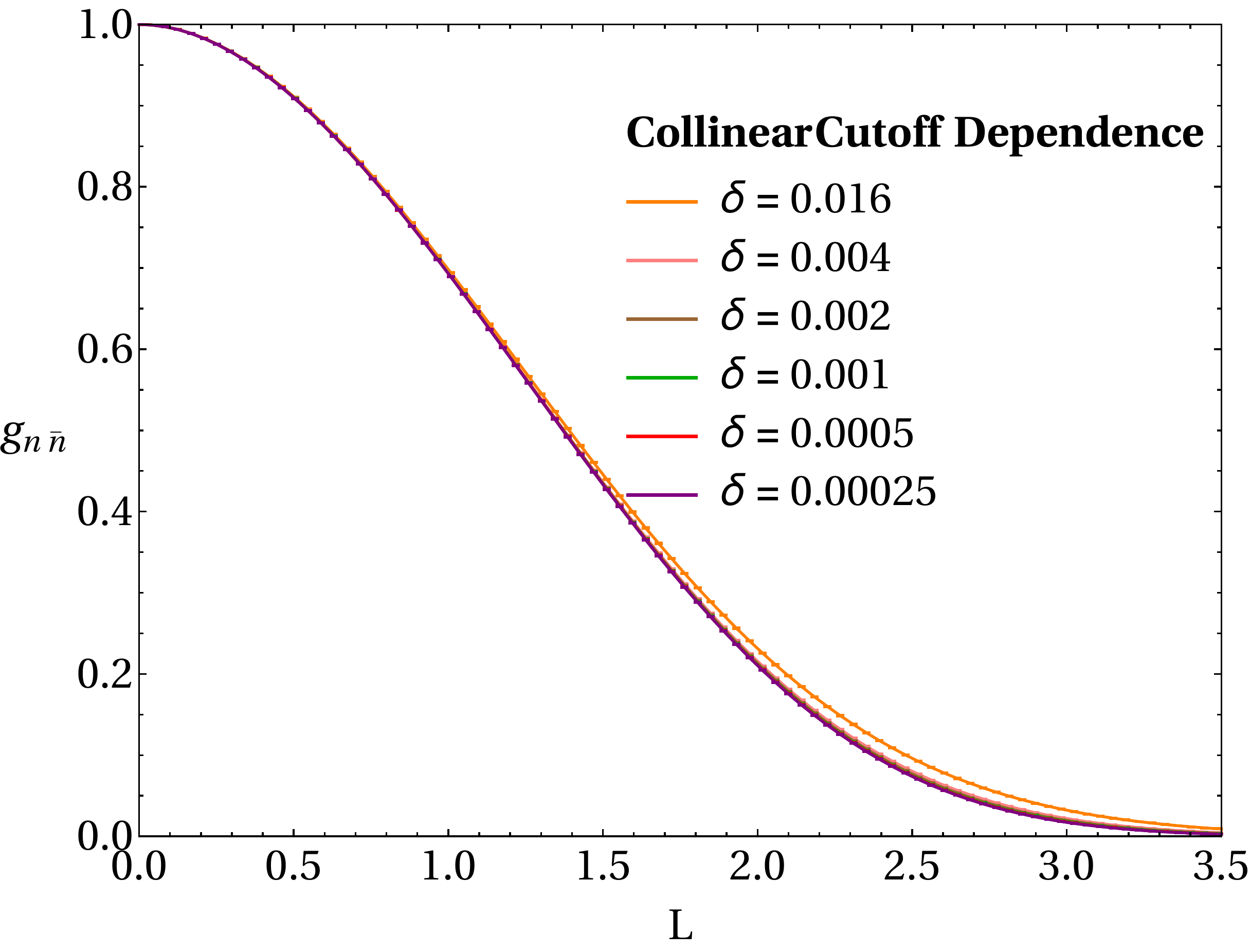}\qquad
\includegraphics[scale=.3]{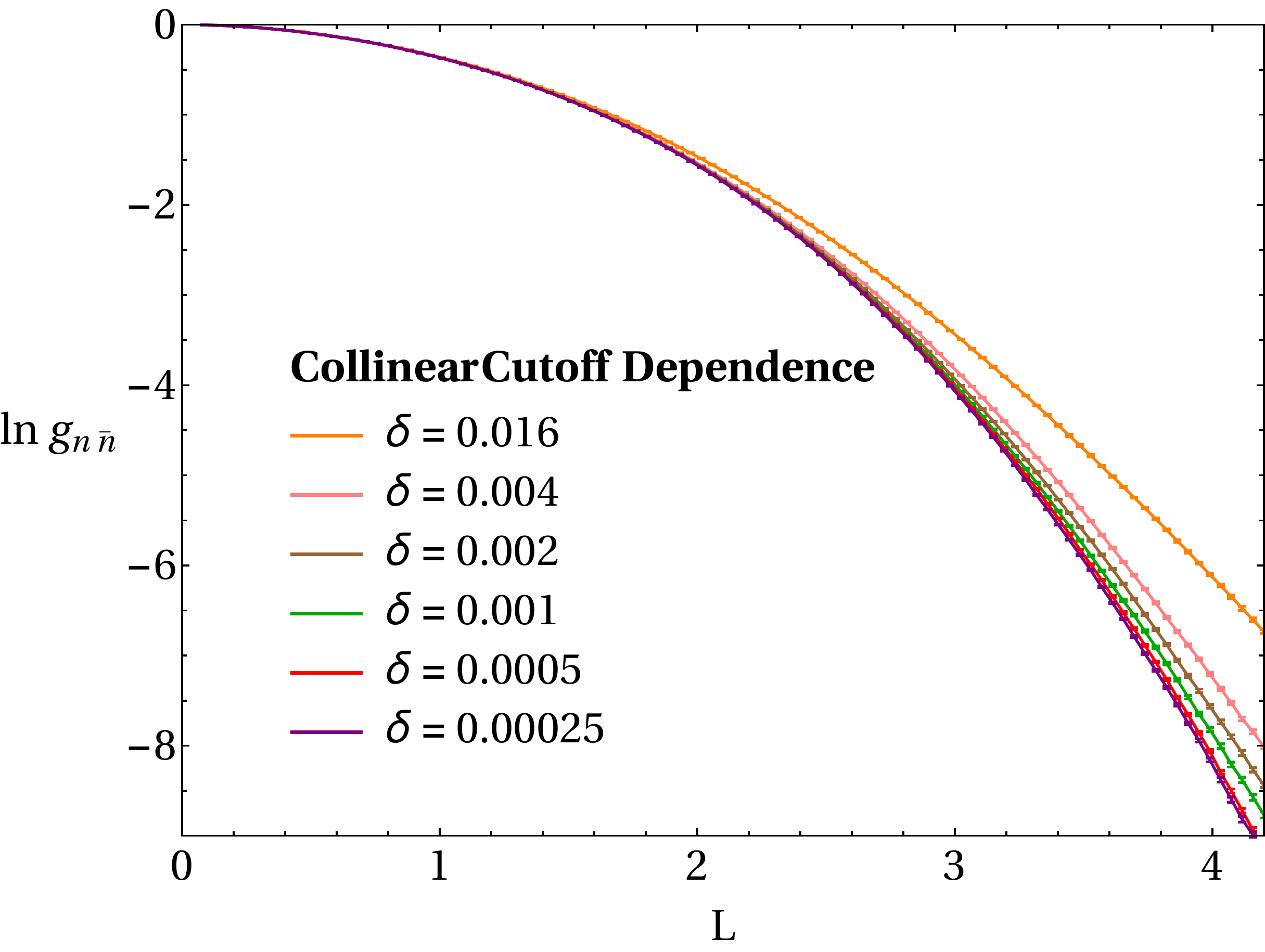}
\caption{\label{fig:decay_constant_MC}The MC data. The MC was run for hemisphere jets with an initial back-to-back dipole configuration.}
\end{figure}
As a check on our derivation, we can use the Monte Carlo (MC) solution of \Ref{Dasgupta:2001sh} to determine the NGL distribution. The Monte Carlo must use an angular cutoff $\delta_{MC}$, which represents the smallest angle any gluon splitting may have with respect to its parent dipole in the initial rest frame of the event. For a description of the determination of the uncertainty for the MC, we refer to \Ref{Larkoski:2016zzc}. We fit the following Gaussian model to the logarithm of the distribution:
\begin{align}\label{eq:NGL_model_tail}
M(L)&=\frac{a}{2}L^2+b\,L\text{ln} L+c L\,.
\end{align}
The fit is only in the tail region, with various lower limits of the fit window taken to be $L=1.5, 2.0,$ and $2.5$. The upper limit $L=4.0$ is where the MC still has reliable error bars for all cutoffs. Under this parameterization, the coefficients $a, b$ should correspond to the coefficients of the leading and subleading terms of the saturation scale in \Eq{eq:saturation_scale}. We map the saturation scale to the collinear cutoff (or the angular size of the boundary of the buffer region) via \Eq{eq:collinear_regulator_to_saturation}. Integrating the logarithm of the cutoff gives the asymptotic behavior parametrized by \Eq{eq:NGL_model_tail}. If we are far enough out in the asymptotic region of the NGL distribution for our fits, as the MC angular cutoff $\delta_{MC}$ is taken to zero, we would hope that the coefficients $a, b$ of the fitted model would obey:
\begin{align} 
\lim_{\delta_{MC}\rightarrow 0}a= -\frac{\chi(\nu_{min})}{2\nu_{min}}\\
\lim_{\delta_{MC}\rightarrow 0}b= \frac{3}{4\nu_{min}}
\end{align}
The first relation is from our derivation of Eqs. \eqref{eq:asymptotic_cutoff} and \eqref{eq:asymptotic_BMS_gen_in-ou}, while the second is relying on the correspondence \Eq{eq:collinear_regulator_to_saturation}. 

\begin{figure}\centering
\includegraphics[scale=.3]{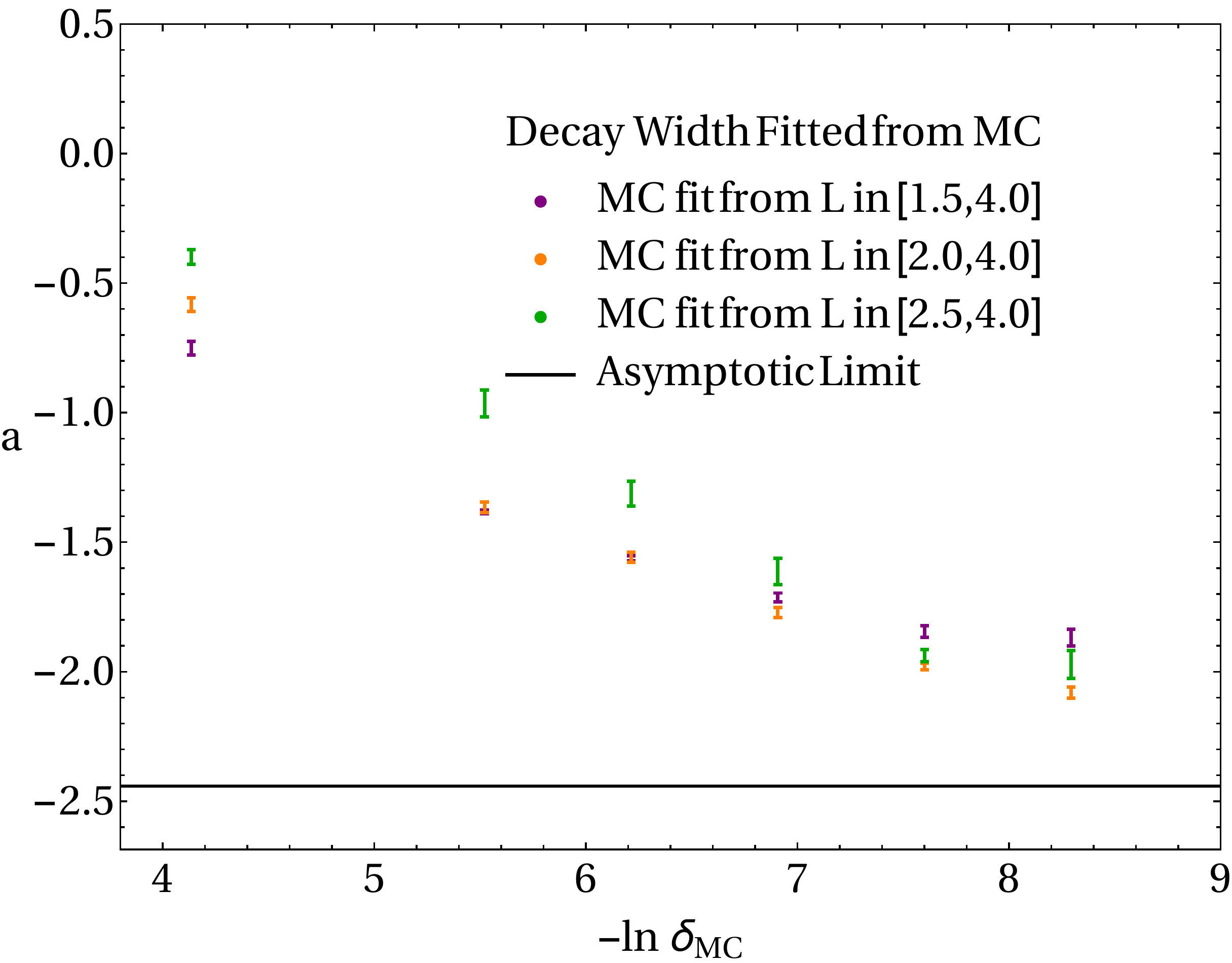}\qquad
\includegraphics[scale=.3]{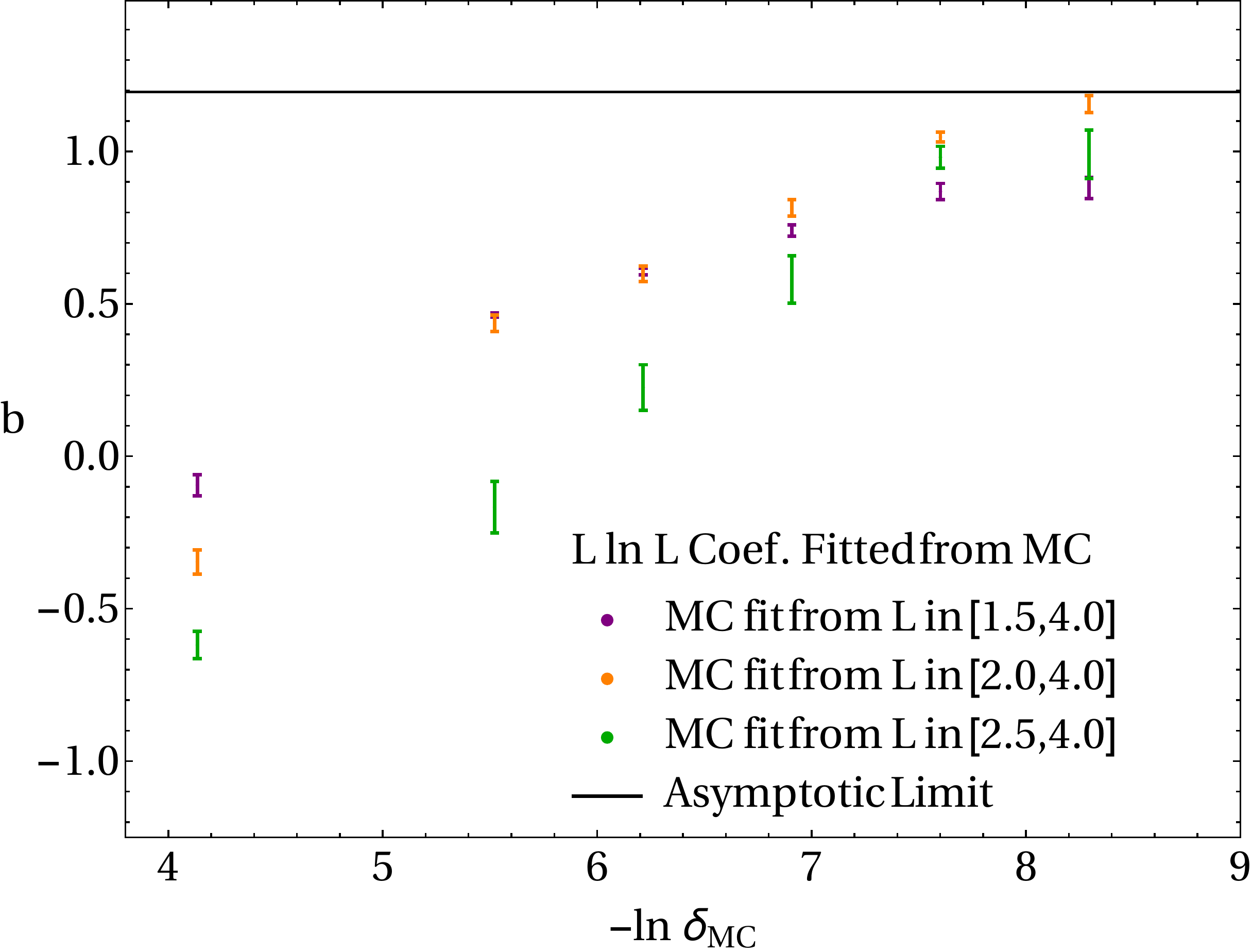}
\caption{\label{fig:LlnL_coef_NGL}The fitted Gaussian decay width and coefficient of the logarithmic term as a function of the MC cutoff. }
\end{figure}

These fits to the tail have a typical $\chi$-squared per degree of freedom very close to 1, improving as the cutoff is lowered. In \Fig{fig:decay_constant_MC} we plot the extracted coefficient $a$ of the $L^2$ term as a function of the MC angular cutoff, and compare to the asymptotic solution in \Eq{eq:asymptotic_cutoff}. The fitted MC points are all above the asymptotic solution, and appear to be approaching the asymptotic result as the cutoff is lowered. In \Fig{fig:LlnL_coef_NGL}, we plot the asymptotic prediction using the relation in \Eq{eq:collinear_regulator_to_saturation} against the MC data, again finding consistency. The fits are sensitive to the fit window, and also are sensitive to the inclusion of another subleading term predicted from saturation in the model for the tail, scaling as the integral of $1/\sqrt{L}$, so of course having more data at lower cutoffs at higher $L$ would be desirable, but not computationally easy. 

\section{Conclusions}
We have given a perturbation theory to extract the asymptotic behavior of the NGL distribution. The basis of the perturbation theory is the collinearly regulated BMS equation, which allows for a linearization with the same initial conditions as the full solution. The solution for this regulated equation is dominated by the collinear poles from the regularization. By choosing a $L$-dependent regularization, we can find the form of the regulator such that we asymptote to the full solution of the BMS equation. This requires careful consideration of the asymptotic solution at the collinear poles of the initial hard jets, where in the full BMS equation we cannot disregard that active emissions can still occur. We expand the BMS equation in this collinear limit, recovering the BFKL equation, which we then solve to find the behavior of the dipoles at large NGL $L$, but small opening angle. Using the BFKL solution at small angles, and demanding that the collinearly regulated solution asymptotes to the full solution via the condition in \Eq{eq:asymptotic_condition}, we find Gaussian behavior for the NGL distribution, and computed the decay width, finding it to be exactly mapped to the black-disc unitarity limit of the BK equation. Moreover, the computed asymptotics at leading order is manifestly independent of the initial jet geometry, since it is wholely determined by the behavior in collinear regions. All that matters is whether we have both initial legs inside or one leg outside the active jet region.

Further, this analysis clarifies the role of the BFKL approximation in the BMS equation. BFKL does \emph{not} correspond to the small $L$ region or dilute region, in contrast to the small-x case. Indeed, the exact initial conditions are known in the non-global evolution, and they are very dilute, having only two active emitters. At small-L jet geometries and the dressing factors of emissions are most important, and the resummation of the non-global logarithms is best understood in terms of the dressed gluon expansion, as argued in \Ref{Larkoski:2016zzc}, where we perturb about progressively more and more dressed asymptotic states. Rather, we must be in an asymptotic regime at large $L$ before we can find BFKL physics emerging in NGLs as a collinear approximation.

What is most remarkable is the similarities between the collinear cutoff and the saturation scale of forward scattering physics. The BK equation can also have Gaussian behavior in its solutions, near the unitary limit, where the asymptotics are controlled by the saturation scale. The analytic form of the saturation scale can be tied to features of fronts of branching random walks. This suggests that the collinear cutoff can be interpreted as a position of a front of a branching random walk, the front being the location of the buffer region measured relative to the hard initial legs of the dipole. From the dressed gluon expansion of \Refs{Larkoski:2015zka,Neill:2015nya,Larkoski:2016zzc}, we know the existence of the buffer region is a universal feature of NGL physics, even at small $L$.

It is worth contrasting the physical pictures developed for the two cases. The NGL distribution receives its most important features from the dynamics of the buffer region, an emergent phenomona where real emissions at the edge of the active jet region are suppressed. At first this suppression is simply a power-law approach to zero dressing each emission as it approaches the jet boundary, as discussed in detail in Refs. \cite{Larkoski:2015zka,Neill:2015nya,Larkoski:2016zzc}. This dressing sets the limit to the validity of fixed order perturbation theory. As the NGL grows, so does the suppression, and in the asymptotic regime real emission becomes \emph{exponentially} suppressed outside a small collinear region around the initial hard leg(s): at large angles out from the hard legs to the edge of the jet one finds a gluonic desert. The angular size of the region around the hard initial legs where active emissions can still occur is given by \Eq{eq:cutoff_from_BFKL}, and this angle gives the boundary of the buffer region as a function of the non-global logarithm. This then defines the boundary of the buffer region at very large NGLs. Saturation, on the other is perhaps the very opposite physical picture, where the nucleus/nucleon becomes an object overstuffed with ``wee'' partons. What perhaps unites the two ideas is that there is a well-defined front in either the transverse plane or the celestial sphere, evolving with either the rapidity or the NGL. 

At higher orders in perturbation theory for the BMS equation, we do not expect the asymptotics to change much, except for the corrections to the BFKL characteristic function, at least in the large-$N_c$ limit. It is reasonable to suppose the higher order nonlinearites are exponentially suppressed relative to the leading order kernel, except in the collinear regimes where they modify the BFKL equation. The $\alpha_s$ corrections to the form of the leading order kernel are governed by the cusp anomalous dimension, see \Refs{Caron-Huot:2015bja,Caron-Huot:2016tzz}. Thus at higher orders, the asymptotic behavior would still be dominated by \Eq{eq:BMS_LL}, but with a simple remapping by what we mean by $L$ relative to the physical measurements $m_H$ and $m_L$:
\begin{align}\label{eq:L_at_higher_orders}
L&=\int_{m_L}^{m_H}\frac{d\mu}{\mu}\Gamma_{\text{cusp}}[\alpha_s(\mu)]
\end{align}
The running coupling does not explicit enter the BMS kernel the same way as it does in the BK kernel, since angles have no intrinsic mass scale associated with them.\footnote{This is not true for emissions at the edge of the jet, \cite{Neill:2015nya,Larkoski:2016zzc}. However, emissions at the edge of the jet we have already shown to be unimportant for the asymptotic distribution.} The mapping between the small-x and the NGL case will begin to break down beyond leading log for a non-conformal theory, so for full QCD, these corrections would be interesting to study. However, one must be careful since the dipole functions themselves $g_{ab}$ also receive matrix element corrections coming from the out-of-active-jet region. Whether this could change the asymptotic behavior would remain to be seen. 

What is perhaps even more important to be investigating is the effect that the full collinear splitting functions would have on the asymptotics at higher orders. These effects are known to be important in the BFKL kernel \cite{Salam:1998tj,Ciafaloni:1999yw,Ciafaloni:2003rd,Vera:2005jt} and BK \cite{Iancu:2015vea,Iancu:2015joa} equations. Though we started working in a manifestly soft regime, collinear splittings at the smallest angles dominate the determination of the distribution. One would want to use the full splitting kernel then to describe the evolution of the dipoles at the smallest angles, matched to the BFKL results.

Since the asymptotic behavior is controlled by the collinear limits of the eikonal factor, it is worth hypothesizing that for the full color leading logarithmic NGL distribution must settle into the same asymptotic behavior due to color coherence.\footnote{For an explicit comparison of the full color NGL distribution to the infinite-$N_c$, see \Ref{Hagiwara:2015bia}.} In the collinear limit, the BFKL equation obtained is only sensitive to the total color charge of that particular leg, but one expects the form of the equation to be identical to the large-$N_c$ case. Thus, we must rescale each leg's contribution by the correct color casimir, and use the full color cusp anomalous dimension in \Eq{eq:L_at_higher_orders}. Also at finite $N_c$, we can have more than two hard legs inside an active jet region coherently evolving. This would be worth investigating further, for it would give a dynamical reason why the large-$N_c$ approximation can work so well in describing the branching history of QCD.

Finally, it would be worthwhile to pursue the explicit calculation of the subleading terms in the asymptotic expansion. The leading behavior is active-jet geometry independent, and it is plausible this remains true for the subleading orders, especially given the universality found in branching random walks. That we could completely recover the saturation scale behavior is already hinted at numerically, since fitting the Gaussian model with the form given in \Eq{fig:decay_constant_MC} was necessary to get the leading coefficient consistent with the asymptotic result, and the subleading logarithmic term was also consistent with the prediction lifted from the saturation scale. However, the subleading asymptotics would require finding the solution more accurately in the intermediate region between the small angle region and the buffer region dominated ansatz, and this may become sensitive to the jet geometry.

\section{Acknowledgments}
I would like to thank Mrinal Dasgupta for discussions of numerical work in determining the form of the tail of the NGL distribution. I would also like to acknowledge discussions with my fellow conspirators on NGLs and subjets, Ian Moult and Andrew Larkoski. I acknowledge support from DOE contract DE-AC52-06NA25396 and through the LANL/LDRD Program. I also thank the Erwin Schrodinger Institute's program ``Challenges and Concepts for Field Theory and Applications in the Era of the LHC Run-2", where portions of this work were completed.

\appendix

\section{ Regulated Eikonal Integral Using The $\delta$-Regulator}\label{sec:eikonal_integrals}
\begin{figure}\centering
\includegraphics[scale=.3]{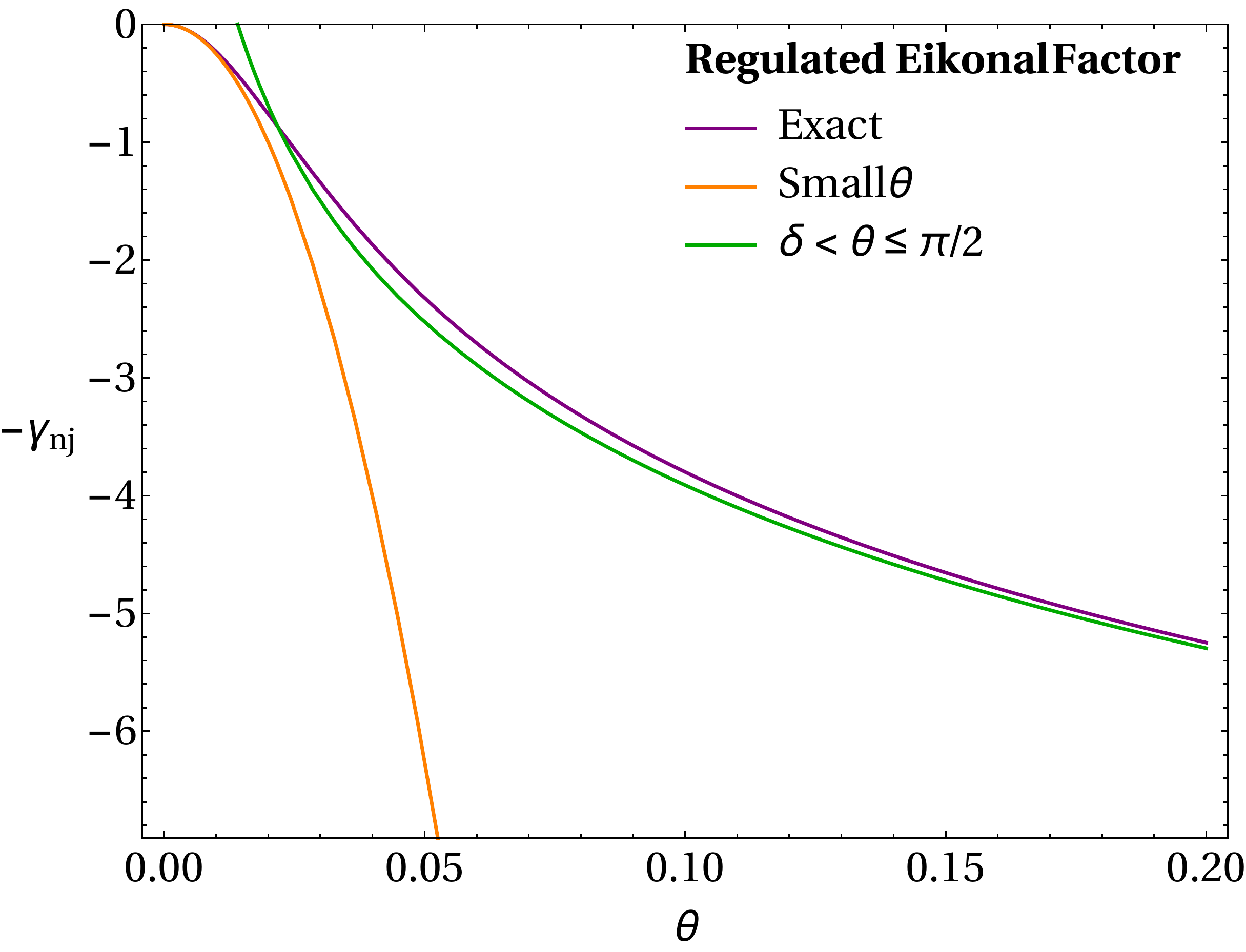}
\includegraphics[scale=.3]{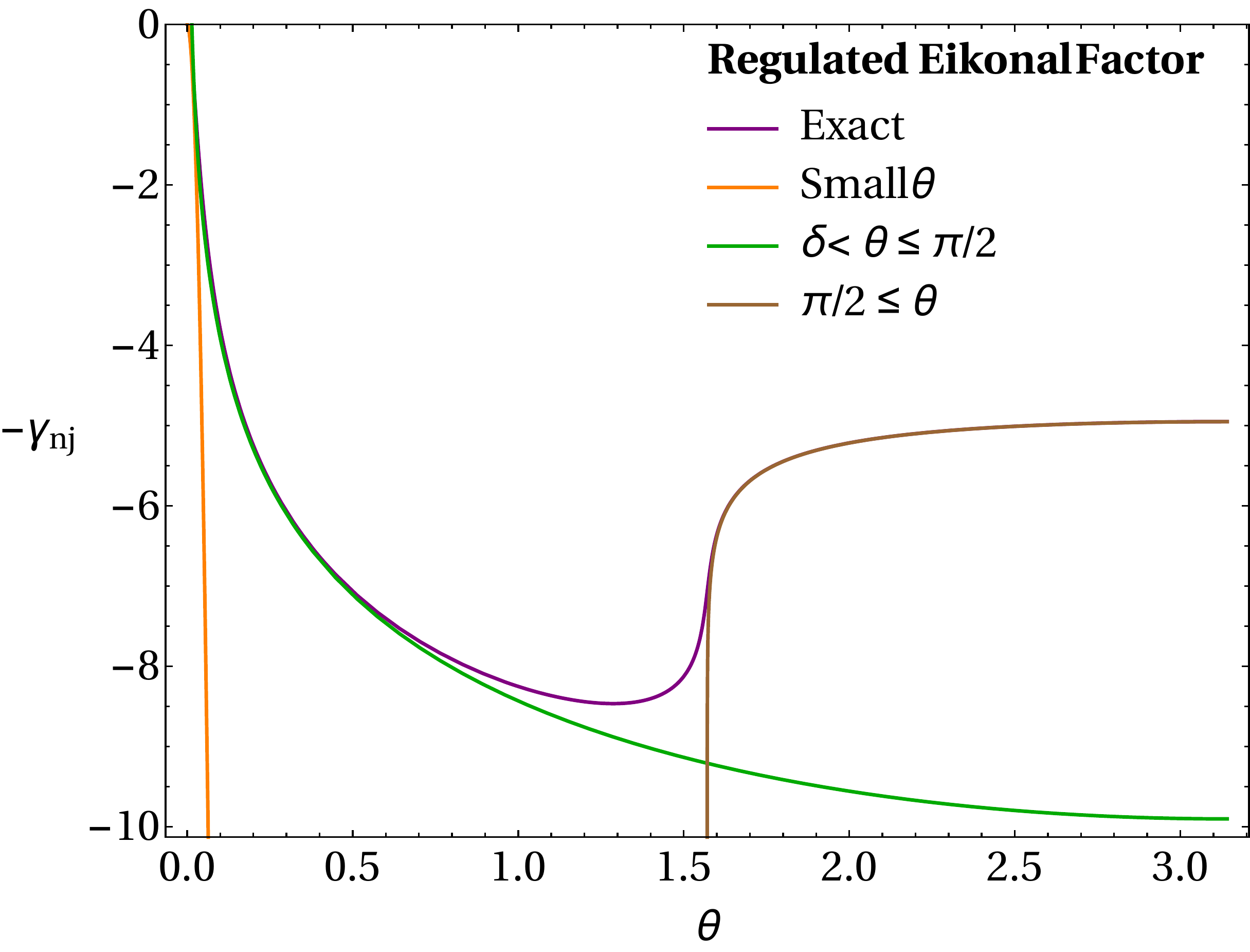}
\caption{\label{fig:reg_eik}The $\delta$-regulated eikonal factor as a function of the angle between the emission and the active jet in a hemisphere dijet geometry. }
\end{figure}
\begin{figure}\centering
\includegraphics[scale=.3]{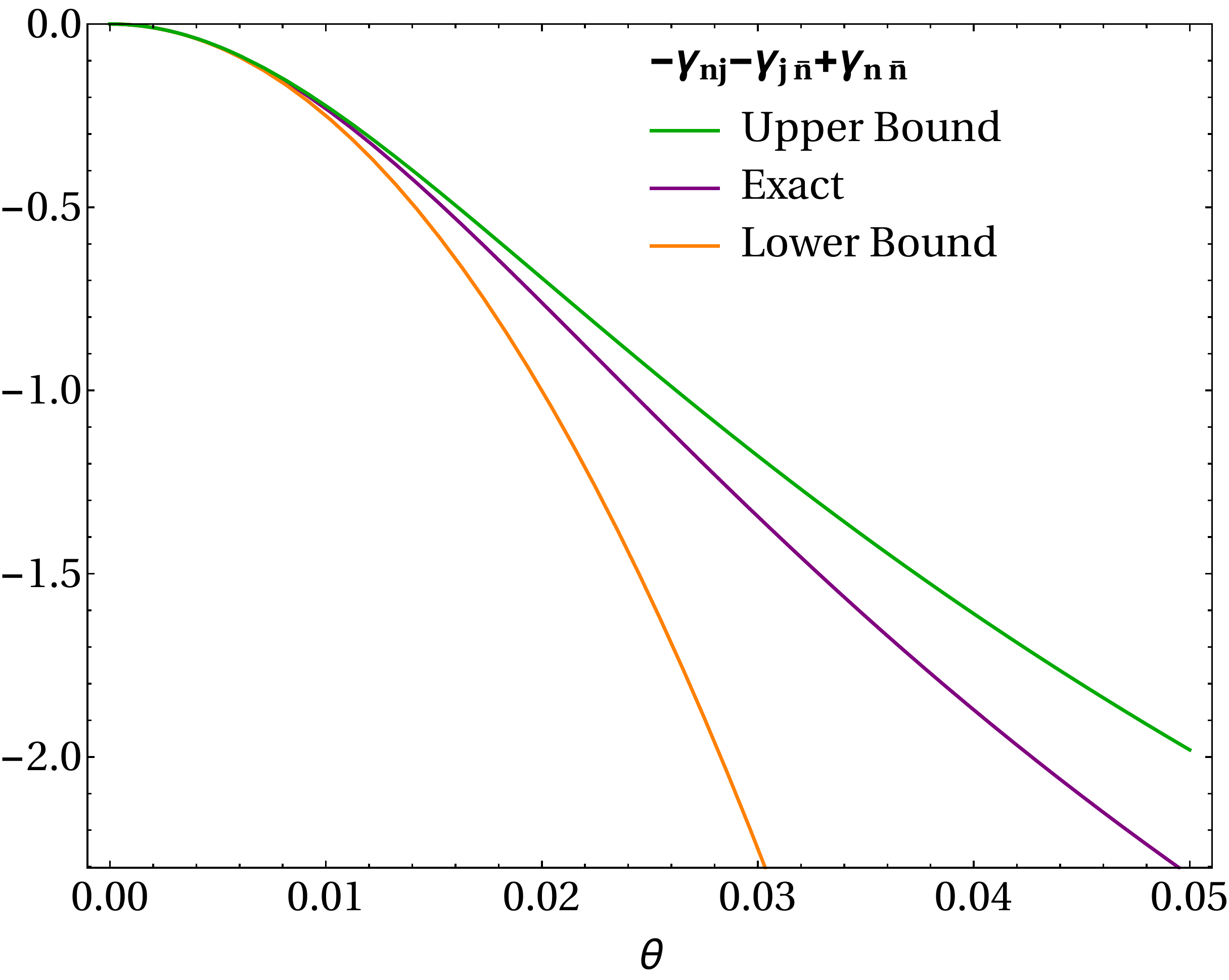}
\includegraphics[scale=.3]{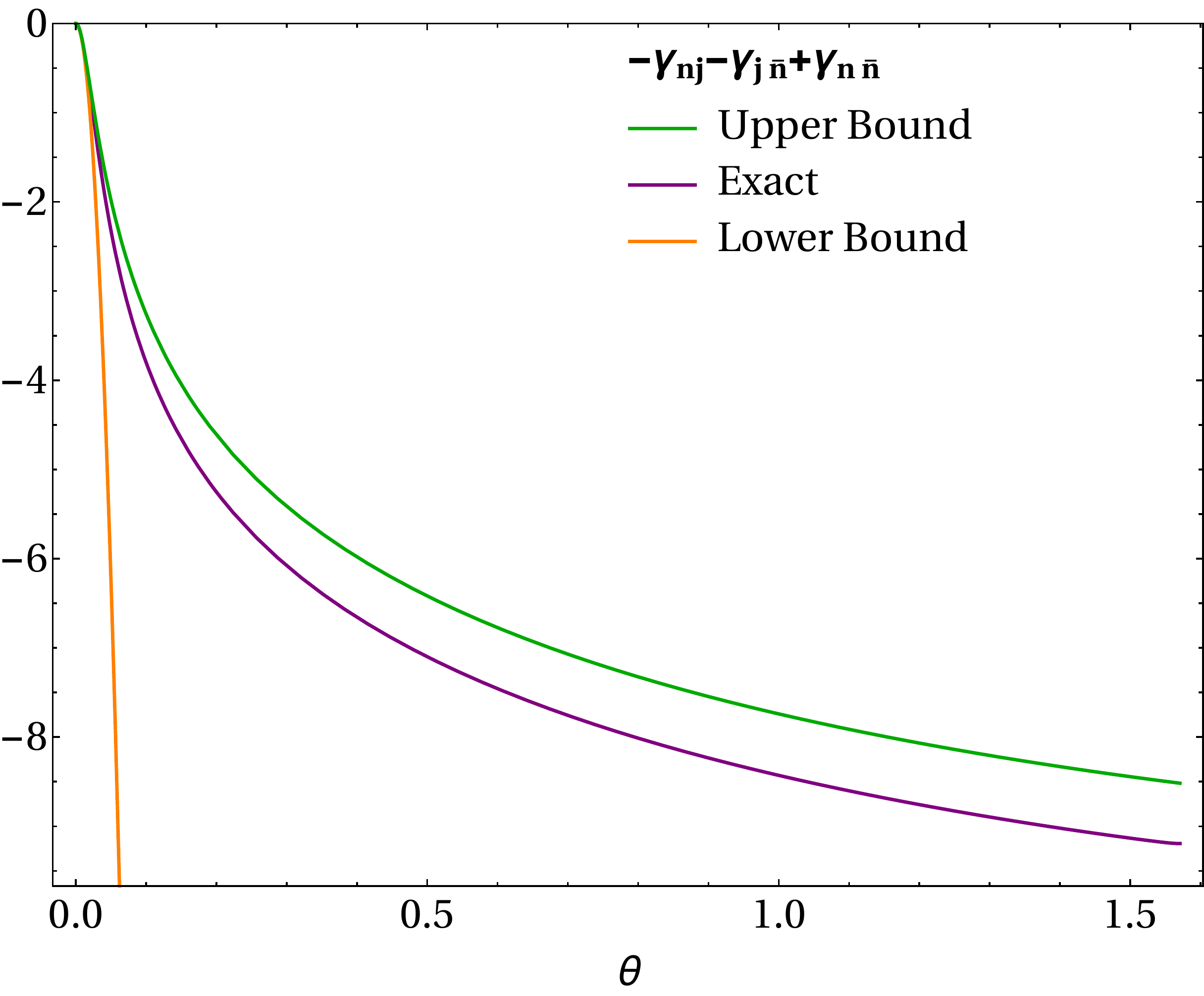}
\caption{\label{fig:exponent_bounds}Bounds on the exponent of the real emission term in the BMS kernel acting on the collinearly regulated ansatz. }
\end{figure}

We evaluate the regulated eikonal integral when the regulator is a small parameter added to the eikonal propagator. We have for an arbitrary jet area $J$:
\begin{align}\label{eq:delta_reg_eikonal}
\gamma_{ab}(\delta,J)&=\int_J\frac{d\Omega_j}{4\pi}W^{\delta}_{ab}(j)=\int_J\frac{d\Omega_j}{4\pi}\frac{1-\hat{a}\cdot\hat{b}}{(1+\delta^2-\hat{a}\cdot\hat{j})(1+\delta^2-\hat{j}\cdot\hat{b})}\,.
\end{align}
We also adopt the notation:
\begin{align}
W_{ab}(j)&=W^{0}_{ab}(j)\,,\\
\theta_{ab}:& \text{ the angle between } \hat{a}\text{ and } \hat{b}\,,\\
D_a^r&=\{x\in S^2: \theta_{ax} \leq r\}\,.
\end{align} 
The regulated eikonal factor is most easily computed when $J=D_a^r$, a conical region of radius $r$ about the direction $r$, and can be done analytically. It's form is not very enlightening nor useful, so we plot the exact result against various approximations in the case $\delta=0.01$ and $r=\frac{\pi}{2}$ in \Fig{fig:reg_eik}. The approximations use in \Fig{fig:reg_eik} are all valid when $\delta\ll 1$, and are given by:
\begin{align}\label{eq:regulated_eikonal_limits}
\gamma_{ab}(\delta,D_a^r)=\begin{cases}\frac{\theta_{ab}^2}{4\delta^2}+...,&\text{ if } \frac{\theta_{ab}^2}{\delta^2}\ll r^2\\
\text{ln}\frac{2\text{sin}^2\frac{\theta_{ab}}{2}}{\delta^2}+...&\text{ if } \delta^2<\theta_{ab}^2<r^2\\
\frac{1}{2}\text{ln}\frac{2\text{sin}^2\frac{\theta_{ab}}{2}\text{sin}^2\frac{R}{2}}{\delta^2\Big(\text{sin}^2\frac{\theta_{ab}}{2}-\text{sin}^2\frac{R}{2}\Big)}+...&\text{ if }\delta^2<r^2<\theta_{ab}^2
\end{cases}
\end{align}
We plot the approximations to the exponent of the exponential term in the BMS kernel acting on the collinear regulated ansatz, found in \Eq{eq:angular_behavior_of_soft_eikonal}, for $r=\frac{\pi}{2}$ and $\delta=.01$ in \Fig{fig:exponent_bounds}. The upper bound is constructed to capture both the small $\theta_{ab}$ dependence, and the angular dependence at large angles of \Eq{eq:regulated_eikonal_limits}. These bounds are accurate in the limit $\delta\ll 1$, and are amiable to analytic analysis. 
We note:
\begin{align}\label{eq:nnbar_gamma}
\gamma_{n\nbar}(\delta,D_n^R)&=\text{ln}\frac{\sqrt{2}\text{tan}\frac{R}{2}}{\delta}+...
\end{align}
It is worth noting that the calculation of the jet region dependence on the ansatz is remarkably similar to that found in one-loop soft calculations for the \emph{global} anomalous dimensions \cite{Ellis:2010rwa,Chien:2015cka,Kolodrubetz:2016dzb}.

\pagebreak
\bibliography{dressedgluon}

\end{document}